\documentclass[aps,twocolumn,floatfix,nofootinbib,preprintnumbers,amsmath,amssymb]{revtex4}

\usepackage{graphicx}
\usepackage{dcolumn}
\usepackage{bm}

\def\beq{\begin{equation}}
\def\eeq{\end{equation}}
\def\bea{\begin{eqnarray}}
\def\eea{\end{eqnarray}}

\newcommand{\lsim}{ \mathop{}_{\textstyle \sim}^{\textstyle <} }

\usepackage{color}

\begin{document}

\preprint{UCI-HEP-TR-2010-09}

\title{Constraints on Light Majorana Dark Matter from Colliders}

\author{Jessica Goodman}
\author{Masahiro Ibe}
\author{Arvind Rajaraman}
\author{William Shepherd}
\author{Tim M.P. Tait}
\author{Hai-Bo Yu}
\affiliation{Department of Physics \& Astronomy, University of California, Irvine, CA 92697}

\date{\today}

\begin{abstract}
We explore model-independent collider constraints on light Majorana dark matter particles.
We find that colliders provide a complementary probe of WIMPs to direct detection,
and give the strongest current constraints on light DM particles. 
Collider experiments  can access interactions not probed by direct detection searches,
and outperform direct detection experiments by about an order of magnitude
 for certain operators in a large part of parameter space.
For operators
which are suppresssed at low momentum transfer,
collider
searches have already placed constraints on such operators limiting their use as an
explanation for DAMA.
\end{abstract}


\maketitle

\section{Introduction}


Recently, there has been much interest in light (order $\sim$~GeV) mass dark matter
\cite{Fitzpatrick:2010em,Kuflik:2010ah,Feldman:2010ke,Chang:2010yk,Essig:2010ye}.
This interest is partly spurred by the fact that the DAMA signal of annual modulation
\cite{Bernabei:2010mq}
may be understood as consistent with null results reported by CDMS \cite{Ahmed:2009zw}
and Xenon 10 \cite{Angle:2007uj} if the dark matter is a weakly interacting massive particle (WIMP)
of mass  $\lesssim 10$~GeV
\cite{Petriello:2008jj}.
Further excitement is
motivated by the signal reported by CoGeNT, which favors a WIMP in the
same mass range \cite{Aalseth:2010vx} as DAMA with moderate channeling (however,
unpublished data from 5 towers of CDMS Si detectors \cite{CDMS-Si}
provides some tension, see \cite{Chang:2010yk}).

A WIMP which is relevant for direct detection experiments necessarily has substantial
coupling to nucleons, and thus can be produced in high energy particle physics experiments
such as the Tevatron and Large Hadron Collider (LHC).  
In particular, light WIMP states can be produced
with very large rates.  
These WIMPs escape undetected, and hence the most promising signals involve missing energy
from a pair of WIMPs recoiling against Standard Model
(SM) radiation from the initial state quarks/gluons
\cite{Birkedal:2004xn,Cao:2009uw,Beltran:2010ww}.  While such searches are complicated 
by large SM backgrounds producing missing energy, 
we will find that colliders can provide stringent
restrictions on the parameter space of light dark matter models.
Colliders can also access interactions which are irrelevant for direct detection
(either because they lead to vanishing matrix elements in non-relativistic nucleon states
or are suppressed at low momentum transfer).  

In this article, we explore the bounds colliders can place on a light
Majorana 
fermion WIMP, which we assume
interacts with the SM largely through higher dimensional operators.
By exploring the complete set of
leading operators, we arrive at a model-independent picture
(up to our assumptions) of WIMP interactions with
SM particles in the case where the WIMP is somewhat lighter than any other particles
in the dark sector. We show that colliders can outperform direct detection 
searches significantly over a large area of parameter space.

\section{The Effective Theory}
\label{sec:model}

We assume that the WIMP ($\chi$)
is the only degree of freedom beyond the SM
accessible to the experiments of interest.
Under this assumption, the interactions between WIMPs and SM fields are mediated by
higher dimensional operators, which
are non-renormalizable in the strict sense, but may remain
predictive with respect to experiments whose energies are low compared to the mass
scale of their coefficients.
We assume the WIMP is a 
SM singlet, and  examine operators of the form \cite{Cao:2009uw,beltran09,Shepherd:2009sa}
\begin{eqnarray}
\label{eq:operators}
{\cal L}_{{\rm int},qq}^{(\rm dim 6)} &=& G_\chi\left[ \bar{\chi} \Gamma^{\chi} \chi\right]
\times \left[ \bar{q} \Gamma^q q \right]\ ,\cr
{\cal L}_{{\rm int},GG}^{(\rm dim 7)} &=& G_\chi\left[ \bar{\chi}\Gamma^{\chi} \chi\right]
\times ( GG\, {\rm or}\, G\tilde{G} )\ ,
\end{eqnarray}
Here  
$q$ denotes the quarks $q=u,d,s,c,b,t$,
and $G$ and $\tilde{G}$ the field strength of the gluon with $\tilde{G}^{\mu\nu}=\epsilon^{\mu\nu\rho\sigma}
G_{\rho\sigma}/2$. %
Ten independent Lorentz-invariant interactions are allowed; 
by applying Fierz transformations, all other
operators can be rewritten as a linear combination
of operators of the desired form.
In Table \,\ref{tab:operators}, we present 
couplings $G_\chi$ and $\Gamma^{\chi, q}$
for these ten operators,
where we have expressed $G_\chi$'s in terms of  an energy scale $M_*$.
\begin{table}[tdp]
\begin{center}
\begin{tabular}{|c|c|c|c|c|}
\hline
   Name    & Type & $G_\chi$ & $\Gamma^\chi$ & $\Gamma^q$ \\
\hline
M1 & $qq$ & $m_q/2M_*^3$  &  $1$                     & $1$ \\
M2 & $qq$ & $im_q/2M_*^3$  &  $\gamma_5$                     & $1$ \\
M3 & $qq$ & $im_q/2M_*^3$  &  $1$                     & $\gamma_5$ \\
M4 & $qq$ & $m_q/2M_*^3$  &  $\gamma_5$                     & $\gamma_5$ \\
M5 & $qq$ & $1/2M_*^2$  &  $\gamma_5\gamma_\mu$                     & $\gamma^{\mu}$ \\
M6 & $qq$ & $1/2M_*^2$  &  $\gamma_5\gamma_\mu$                     & $\gamma_5\gamma^{\mu}$ \\
M7 & $GG$ & $\alpha_s/8M_*^3$  &  $1$                     & - \\
M8 & $GG$ & $i\alpha_s/8M_*^3$  &  $\gamma_5$                     & - \\
M9 & $G\tilde{G}$ & $\alpha_s/8M_*^3$  &  $1$                     & - \\
M10 & $G\tilde{G}$ & $i\alpha_s/8M_*^3$  &  $\gamma_5$                     & - \\
\hline
\end{tabular}
\caption{The list of the effective operators defined in Eq.\,(\ref{eq:operators}).
\label{tab:operators}}
\end{center}
\end{table}%
In the table, we have assumed that the coefficients of the scalar operators, M1-M4, are
proportional to the quark masses, in order
to avoid large flavor changing
neutral currents.
We will assume that the interaction is dominated by only one of the above operators
in the table.


Our effective theory description will break down at energies of order
the mass of whatever virtual particles mediate the $\chi$ - SM interactions.  If we imagine that
the interactions are mediated at tree-level by some heavy state with couplings of order $g$ and
mass $M$, we can identify $M_* \sim M / g$.  In order for perturbation theory to be trustworthy,
$g \lsim 2 \pi$, and thus the effective theory description can at best be valid for
$M_* \geq m_\chi / (2 \pi)$, providing an upper limit on collider cross sections for which there
can be any effective theory description.  

\begin{figure}[t]
\includegraphics[width=8.0cm]{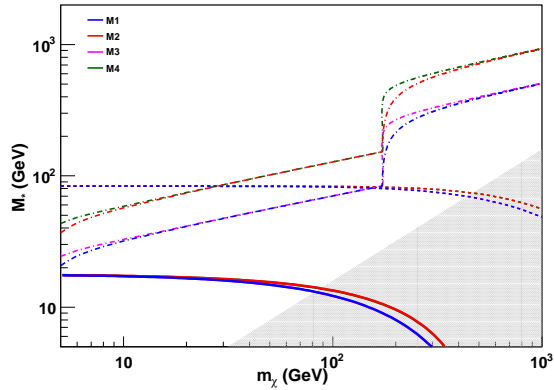}
\caption{\label{fig:1-4} 
Constraints on $M_*$ for  operators M1-M4.
Solid lines are Tevatron 2$\sigma$ constraints.
Dashed lines show LHC 5$\sigma$ reach.
Results for M1 and M2 are largely degenerate with  M3 and M4, respectively.
The dash-dotted lines show the value of $M_*$ 
which reproduce the thermal relic density ($\Omega h^2 = 0.11$).
In the shaded region, the effective theory breaks down.
}
\end{figure}

\section{Collider Constraints}

We put constraints on each operator in Table\,\ref{tab:operators}
by considering 
 the pair production of  WIMPs at hadron colliders together with associated hard jets,
\begin{eqnarray}
 p\bar{p}\, (pp) \to \chi\chi + {\rm jets}\ .
\end{eqnarray}
We generate  signal events  for each operator 
using Comphep\,\cite{Pukhov:1999gg,Boos:2004kh}
and shower them with Pythia\,\cite{Sjostrand:2006za}
with the help of the Comphep-Pythia interface\,\cite{Belyaev:2000wn}.
Detector effects are simulated using PGS\,\cite{PGS} with the CDF
detector model.

The largest and irreducible Standard Model background is $Z+$jets with $Z\rightarrow \nu\nu$.
The next important  background comes from $W$+~jets where the charged lepton from $W$-decay 
is lost.  The QCD multi-jet production with mismeasured transverse momentum also contributes to 
the background, but is expected to be subdominant for our 
cuts \cite{Alwall:2008va,Aaltonen:2008hh}.

At the Tevatron, monojet searches~\,\cite{Aaltonen:2008hh,CDF} have looked for events with
leading jet $E_T>80$\,GeV, missing $E_T>80$\,GeV, 2nd jet with $p_T<30$\,GeV, and vetoing any
3rd jet with $E_T>20$\,GeV.
CDF analyzed 
$1.0$~fb$^{-1}$ of data 
with $8449$ observed events.
The expected number of SM background events is $ N_{\rm SM} = 8663 \pm 332$.
Based on this result, we put a $2\sigma$ upper limit on the new physics cross section
of $  \sigma_{\rm new} < 0.664\,{\rm pb}$ (after cuts),
which we translate into bounds on $M_*$.

For the LHC, we  simulate jets + missing energy events (without vetoing extra jet activity)
for $\sqrt{s} = 14$\,TeV
and compare them with the analysis in Ref.\,\cite{Vacavant:2001sd}.
In Ref.\,\cite{Vacavant:2001sd}, the number of 
SM background events
with missing $p_T$ larger than $500$\,GeV
was about $B=2\times10^4$  for integrated luminosity $100$\,fb$^{-1}$, while
the signal acceptance is better than 90\%.  We
assume that the signal acceptance remains 90\%, that is,
$ S  = 0.9 \times \sigma_{j\chi\chi} \times  100\,{\rm fb}^{-1}$,
where $\sigma_{j\chi\chi}$ is the parton-level cross section.
We define the $5\sigma$ reach at the LHC by
$ S / \sqrt{B} >5$, which we translate into reaches on $M_*$.

The Tevatron constraints and LHC reaches on $M_*$ for each operator in 
Table\,\ref{tab:operators}  are summarized in Figs 1-3. These bounds on $M_*$ 
can be applied generically to models of of dark matter, and can be used to place constraints. 
In the following, 
we apply them to find new constraints on direct detection cross sections.


\begin{figure}[t]
\includegraphics[width=8.0cm]{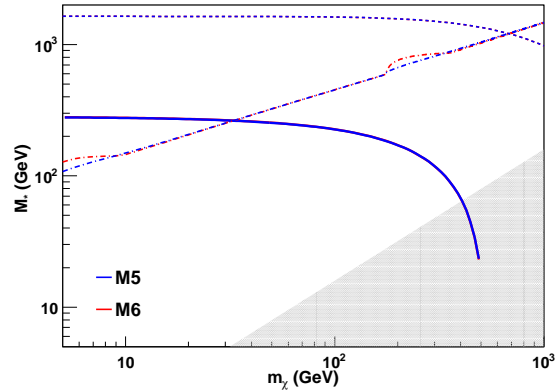}
\caption{\label{fig:5-6} 
Same as Fig.\,\ref{fig:1-4}, but for the largely degenerate operators M5 and M6.
}
\end{figure}

\begin{figure}[t]
\includegraphics[width=8.0cm]{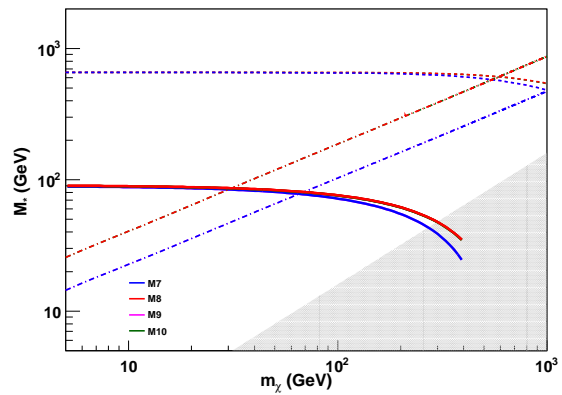}
\caption{\label{fig:7-10} 
Same as Fig.\,\ref{fig:1-4}, but for the operators M7 and M8 which are largely
degenerate with M9 and M10, respectively.
}
\end{figure}

\begin{figure}[t]
\includegraphics[width=9.5cm]{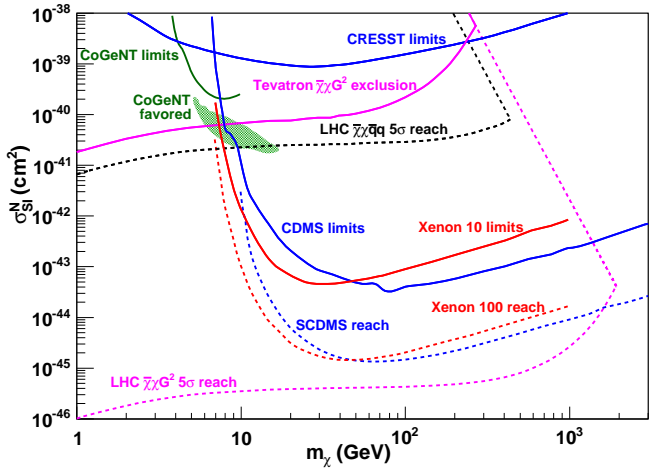}
\caption{\label{fig:si-msigma} 
Regions of parameter space 
excluded by Tevatron searches,
CDMS/Xenon 10\,\cite{Ahmed:2009zw,Angle:2007uj}, CoGeNT\,\cite{Aalseth:2008}, and CRESST\,\cite{Angloher:2002in}  (solid lines as indicated).  The shaded
region is the parameter space favored by a WIMP interpretation of the CoGeNT signal\,\cite{Aalseth:2010vx}.
Also shown are projected bounds for 
for the LHC, (S)CDMS\,\cite{Akerib:2006}, and Xenon 100\,\cite{Aprile:2009yh} 
(dotted lines as indicated).  
}
\end{figure}

\begin{figure}[t]
\includegraphics[width=9.5cm]{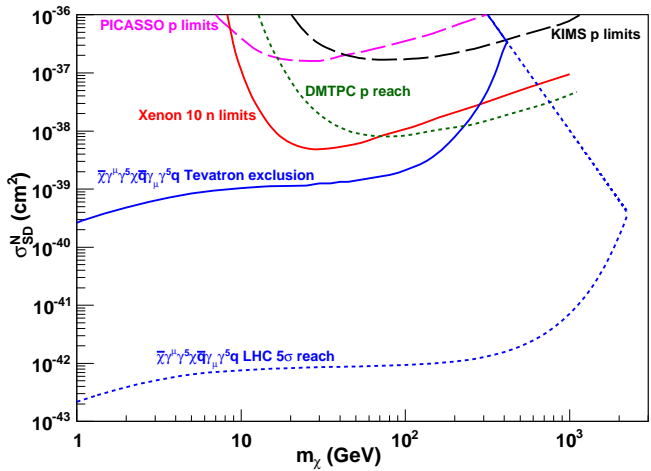}
\caption{\label{fig:sd-msigma} Regions of parameter space 
excluded by Tevatron searches,
Xenon 10\,\cite{Angle:2007uj}, KIMS\,\cite{Lee:2007} and PICASSO\,\cite{Archambault:2009}.
Also shown are projected bounds
for the LHC and DMTPC\,\cite{Sciolla:2009}.  }
\end{figure}

\begin{figure}[t]
\includegraphics[width=9.5cm]{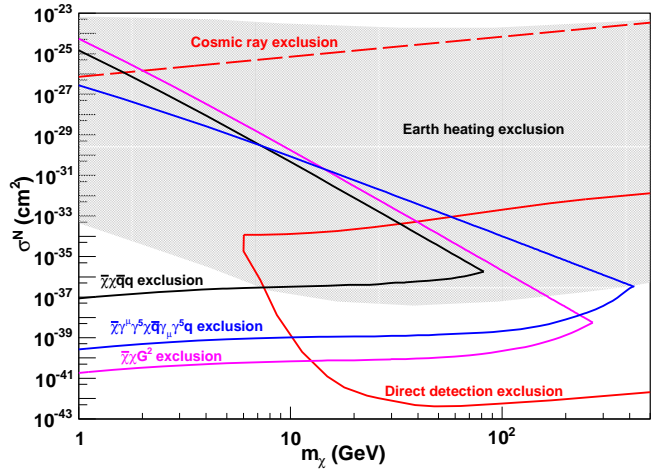}
\caption{\label{fig:zoom} The regions of parameter space 
excluded by Tevatron and other constraints (taken from \cite{Mack:2007}).
 }
\end{figure}

\section{Implications for direct detection}
Only operators M1, M6, and M7 contribute to direct detection in the limit of zero
momentum transfer. Through standard calculations 
\,\cite{Belanger:2008}
we find that the single-nucleon cross sections due to these operators are
\begin{eqnarray}
\sigma_{SD;M6}^{N} &=& \frac{16\mu_\chi^2}{\pi}\left(0.015\right)\left(\frac{1}{2M_*^2}\right)^2 ,\\
\sigma_{SI;M1}^{N} &=& \frac{4\mu_\chi^2}{\pi}\left(0.082\ {\rm GeV}^2\right)\left(\frac{1}{2M_*^3}\right)^2 , \\
\sigma_{SI;M7}^{N} &=& \frac{4\mu_\chi^2}{\pi}\left(5.0\ {\rm GeV}^2\right)\left(\frac{1}{8M_*^3}\right)^2,
\end{eqnarray}
where $\mu_\chi$ is the reduced mass.
We translate our limits on $M_*$ for each operator into a constraint on
the direct detection cross section (for the relevant operators)
which can be induced by that operator. In Figs 4-6, we plot the constraints
from the Tevatron and the discovery reach of the LHC
on the cross sections, as well as other existing constraints.

The most striking feature of our collider-derived constraints 
is the
fact that they are sensitive to arbitrarily light DM particles. They are thus highly complementary
to direct detection experiments, which have limited sensitivity to light DM due to
their finite energy threshold.  For light Majorana WIMPs, colliders make definite and important
statements about the properties of DM.  More generally, models with very low
WIMP masses are most efficiently probed at colliders.

For WIMPs of mass less than 10 GeV, the Tevatron constraints
already rule out cross sections above $\sim 10^{-37}$ cm$^{-2}$, 
which are allowed by all other constraints.
If the DM couples through an operator like $\chi\chi G^2$, the LHC will be able to place bounds far
superior to any near-future DM experiment searching for spin-independent scattering, even for DM 
masses up to a TeV. Spin-dependent experiments are already outperformed in much of parameter 
space by
current Tevatron bounds, while the LHC can place bounds several orders of magnitude better 
than near-future spin-dependent experiments. 

\section{Conclusions}

We have derived new constraints on generic Majorana DM models 
based on null search results for monojets  at the Tevatron,
and explored corresponding discovery reaches at the LHC. Our bounds
cover regions of parameter space which were previously not constrained by experimental efforts
and strongly constrain some kinds of low mass WIMPs as an explanation for the DAMA and
CoGeNT signals. In particular, we have derived constraints on the direct detection scattering of
light Majorana WIMPs which are significantly stronger than experimental bounds (and 
near-future prospects) for spin-dependent scattering.

Colliders are particularly good experiments for testing DM models which are suppressed at 
small momentum transfer, whether the suppression is kinematic in nature as in 
models of light DM, or if the momentum dependence is inherent in 
the induced operator itself, as is the case with momentum-dependent DM\,\cite{Chang:2009}. It would be interesting to study the collider constraints on these models in detail.\\


\noindent
{\em Note added:}  During the final stages of preparing this manuscript, the Xenon 100
collaboration released data constraining the low WIMP mass
region of parameter space \cite{Aprile:2010um}.

\section*{Acknowledgements}
T.T. is glad to acknowledge earlier
collaboration with M.~Beltran, D.~Hooper, E.W.~Kolb,
and Z.~Krusberg and conversations with Chacko, J.Hewett, M. Peskin, and J. Wacker.
This research is supported in part by NSF Grant No. PHY-0653656
and PHY--0709742.

\end{document}